\documentclass[preprintnumbers,10pt,nofootinbib]{revtex4}

\usepackage{amsmath,latexsym,amssymb,amsfonts}
\usepackage{graphicx}
\usepackage{bm}
\usepackage{epstopdf}
\usepackage{epsfig}
\usepackage{color}

\begin{document}

\title{\textbf{Testing black hole equatorial reflection symmetry using Sgr A* shadow images}}

\author{
{\textsc{Che-Yu Chen$^{a}$}\footnote{{\tt b97202056{}@{}gmail.com}}}
}

\affiliation{
$^{a}$\small{Institute of Physics, Academia Sinica, Taipei 11529, Taiwan}
}

\begin{abstract}
The recently released images of the supermassive black holes in the M87 galaxy and the galaxy of our own make probing black hole spacetimes and testing general relativity (GR) possible. The violation of equatorial reflection symmetry of black hole spacetimes is clearly a smoking gun of physics beyond GR. In this paper, we place constraints on the violation of black hole reflection symmetry using the bounds on the ring size of the Sgr A* black hole images, which are consistent with the critical curve radius predicted in GR within $\sim10\%$. Adopting a theory-agnostic framework, we consider a Kerr-like metric in which the violation of reflection symmetry is parametrized by a single parameter, without altering the multipoles of the black hole. The critical curves are always vertically symmetric due to the existence of hidden symmetry associated with the Killing tensor, even though the spacetime is reflection asymmetric. We find that the size of critical curves is sensitive to the amount of reflection symmetry being violated, and place the first constraints on the parameter space of the model using the Sgr A* images. 
 \end{abstract}

\maketitle

\newpage

\tableofcontents

\section{Introduction}

Einstein's general relativity (GR), since its birth in 1915, has kept showing its merits that a lot of theoretical predictions made by GR have been confirmed to be true during this century. Two most recent and important examples are the successive direct detection of gravitational waves emitted by the merger events of binary systems \cite{Abbott:2016blz,LIGOScientific:2018mvr,LIGOScientific:2020ibl}, as well as the images of the supermassive black holes M87* and Sgr A*, respectively, at the center of M87 galaxy \cite{Akiyama:2019cqa,Akiyama:2019brx,Akiyama:2019sww,Akiyama:2019bqs,Akiyama:2019fyp,Akiyama:2019eap} and our own Milky Way \cite{EventHorizonTelescope:2022xnr,EventHorizonTelescope:2022vjs,EventHorizonTelescope:2022wok,EventHorizonTelescope:2022exc,EventHorizonTelescope:2022urf,EventHorizonTelescope:2022xqj}. The most intriguing consequence of these achievements is that we are now certainly able to probe the physics in the spacetime regions with extremely strong gravitational fields such as in the region very close to the black hole horizon, where theoretically some quantum corrections are likely to appear. Although so far we have not found any promising evidence hinting for deviations from GR at the horizon scale, with the developments of the future observational technology, we would be able to capture, if any, those tiny \textit{quantum signals} and if it comes true, that would be extremely exciting.

In the lack of a complete description of the quantum theory of gravity, there are a plethora of ways and models trying to describe the gravitational phenomena in the presence of quantum effects. For example, some effective descriptions of quantum gravity in the low-energy limits, whose quantum correction terms are added to the GR action perturbatively \cite{Endlich:2017tqa,Cardoso:2018ptl,Cano:2019ore}, may modify the black hole geometries at the regions very close to the horizon. Also, from a more phenomenological perspective, one could adopt a theory-agnostic approach, focusing on the phenomenological deviations and introducing the parameters that can parametrize these deviations. More precisely, one can construct artificial black hole metrics (for example, a Kerr-like black hole) which deviate from their GR counterpart, and these deviations are parametrized by some additional parameters in the metric. Very similar to the well-known parameterized post-Newtonian approach in which the deviations from Newtonian gravity are described by some parameters, by constraining the parameters that parametrize the deviations in the Kerr-like metric, we are able to test these black hole solutions and to see whether the no-hair theorem is violated or not. It should be emphasized that the Kerr-like spacetimes are not necessarily solutions to any theories of gravity. They can be constructed arbitrarily, as long as they are able to quantify the deviations from the Kerr metric of interest, and they reduce to the Kerr spacetime in the proper limits. In the literature, the theory-agnostic approach has been used widely in black hole physics and black hole phenomenology \cite{Glampedakis:2005cf,Vigeland:2011ji,Johannsen:2011dh,Johannsen:2015pca,Cardoso:2014rha,Konoplya:2016jvv,Konoplya:2016pmh,Ghasemi-Nodehi:2016wao,Konoplya:2018arm,Bambi:2019tjh,Chen:2019jbs,Konoplya:2020hyk,Carson:2020dez,Chen:2020aix,Delaporte:2022acp}.

In this paper, we consider the theory-agnostic approach and focus on the Kerr-like black holes whose equatorial reflection symmetry, or $\mathbb{Z}_2$ symmetry, is generically broken \cite{Chen:2020aix}. The violation of $\mathbb{Z}_2$ symmetry implies that the physics in the upper hemisphere (with respect to the equatorial plane of the black hole) is different from that in the lower hemisphere. In fact, black hole solutions whose $\mathbb{Z}_2$ symmetry is broken could naturally appear in gravitational theories which contain parity-violating terms. The reflection asymmetry also appears in some string-inspired models \cite{Bah:2021jno}. In the effective field theory of quantum gravity, a series of correction terms are added into the Einstein-Hilbert action. These terms may contain higher order curvature contributions or appear in the form of non-minimal matter-geometry couplings. Essentially, the parity-violating corrections naturally appear when there are higher order curvature terms consisting of the dual Riemann tensor
\begin{equation}
\tilde R_{\mu\nu\alpha\beta}\equiv\frac{1}{2}\epsilon_{\mu\nu\rho\sigma}{R^{\rho\sigma}}_{\alpha\beta}\,,
\end{equation}
and these terms are not zero when black holes are spinning. For example, black hole solutions without $\mathbb{Z}_2$ symmetry are found in Refs.~\cite{Cano:2019ore} and \cite{Cardoso:2018ptl}, where the former is based on a theory with a coupling between the Chern-Simons term and the Gauss-Bonnet term, and the latter corresponds to an effective field theory containing explicit parity-violating curvature invariants. Usually, the rotating black hole solutions in these types of theories are extremely complicated and can only be derived perturbatively, rendering the investigations of such solutions tremendously difficult. In this paper, we consider a class of Kerr-like black holes whose $\mathbb{Z}_2$ asymmetry is parametrized by a deviation parameter $\epsilon$. The metric was proposed in our previous work \cite{Chen:2020aix} and it was found that the deviation parameter $\epsilon$ could alter the size of the shadow, or more precisely, the size of the critical curve in the image. However, in that paper, we did not consider the theoretical requirement that the $\mathbb{Z}_2$ asymmetry is supposed to be induced by the spin of the black hole. As we have mentioned, the reflection asymmetry is usually tightly related to the parity-violating terms in the theory, which are quiescent if the black hole is not spinning. Therefore, it would be more reasonable and eligible to require that the $\mathbb{Z}_2$ asymmetry is induced from the spin of the black hole. In this paper, the violation of such symmetry is designed to be induced by black hole spin. If there is no spin, the $\mathbb{Z}_2$ symmetry is preserved and the Kerr-like metric reduces precisely to the Schwarzschild spacetime. Furthermore, the Kerr-like spacetime is asymptotically flat and it reduces to the Kerr metric at a far distance from the black hole. Also, in order to focus mainly on the effects coming from the reflection asymmetry, the metric of the Kerr-like spacetime will be constructed with the requirement that there exists a hidden symmetry in the spacetime, which is characterized by the Killing tensor. More explicitly, the Kerr-like spacetime is designed to contain a Carter-like constant \cite{Carter:1968ks} and the geodesic equations are completely separable. The simultaneous violation of reflection symmetry and the symmetry associated with the Carter-like constant could generate several interesting signatures, for example, in the image cast by the black hole \cite{Eichhorn:2021etc,Eichhorn:2021iwq}, although, it is generically not straightforward to tell which parts of the signatures result precisely from the reflection asymmetry of the spacetime.  

After constructing the Kerr-like metric with spin-induced $\mathbb{Z}_2$ asymmetry, we will show how the critical curve in the image of the black hole is altered in the presence of the $\mathbb{Z}_2$ asymmetry. Essentially, the critical curve in the image is defined by the impact parameter of the photon region around the black hole \cite{Bardeen1973,Cunha:2018acu}. The photon region of a rotating black hole consists of a number of photon trajectories on which the photons undergo spherical motions. These spherical motions are unstable against radial perturbations. Any tiny perturbations acting on the photons moving on these trajectories would either make the photons fall into the black hole or escape from the potential well then head for spatial infinity. In GR, the shadow critical curve is solely determined by the black hole mass, the spin, and the inclination angle between the line of sight of the observer and the rotation axis of the black hole. If the black hole metric contains additional parameters, as what we will consider here, the shadow critical curve may be different from the GR counterpart even though the two kinds of black holes share the same mass and spin. Alternatively, given an accurate measurement of the size of the image, combined with independent measurements of black hole's mass-to-distance ratio, one is able to constrain the non-GR parameters that are sensitive to the size of critical curves. Note that the inclination angle and the black hole spin are insensitive to the critical curve size.

The idea of using black hole shadow \cite{Cunha:2018acu} to test black hole spacetimes as well as gravitational theories has been proposed for decades \cite{Falcke:1999pj}. However, this idea was not realized until the very recent observation of the image of the supermassive black hole at the center of M87 galaxy. More excitingly, the latest observations of the Sgr A* image would allow us to place tighter constraints using the size measurement. Because the mass-to-distance ratio has been independently measured very accurately using stellar dynamics near the center of our own galaxy. In average, the  measured ring size of the Sgr A* is consistent with the shadow critical curve predicted in GR within $10\%$ \cite{EventHorizonTelescope:2022xqj}. This allows us to test GR and place constraints on deviations from GR, in particular, on the reflection asymmetry of the spacetime. We would also like to mention that the phenomenology of reflection asymmetry of black holes is still largely unexplored, in spite of some previous works that discuss the properties of accretion disk \cite{Datta:2020axm,Chen:2021ryb} and the detectability of using future space-based gravitational wave interferometers \cite{Fransen:2022jtw}.

The paper is organized as follows: In Section~\ref{sectII}, we review the construction of the Kerr-like spacetime in which the $\mathbb{Z}_2$ symmetry is generically broken. In Section~\ref{secIII}, we review the null geodesic equations of the Kerr-like metric and the equations that describe the shadow critical curves. In Section~\ref{sec.IV}, we constrain the reflection asymmetry using the latest ring size measurements of the Sgr A* images. We finally conclude in Section \ref{conclu}.

\section{Rotating black holes without $\mathbb{Z}_2$ symmetry}\label{sectII}
In this section, we will briefly review the model of Ref.~\cite{Chen:2020aix}, which consists of a class of spacetime metrics that can describe a rotating Kerr-like black hole without $\mathbb{Z}_2$ symmetry. We will assume that the spacetime is asymptotically flat. More explicitly, it is required to recover the Kerr metric at a large distance away from the black hole. In addition, we will require that the designed spacetime has a Carter-like constant, such that the geodesic equations defined in the spacetime are completely separable. It should be mentioned that although the second assumption is not necessarily satisfied for the rotating black holes derived from a perturbative approach in effective theories, it can largely simplify the analysis of the geodesic equations and allow us to focus on the effects coming from the reflection asymmetry. From a theory-agnostic point of view, the Kerr-like metric that we are going to construct can not only capture the physical properties of our interest, that is, the violation of $\mathbb{Z}_2$ symmetry, but also retain its astrophysical relevance due to its physical resemblance to the Kerr spacetime in the proper limits. 

The starting point is to consider a general axisymmetric metric that preserves the existence of a Carter-like constant. In Ref.~\cite{Papadopoulos:2018nvd}, the authors proposed a general axisymmetric spacetime (the PK metric) based on this assumption and the geodesic equations of the PK metric are separable. Very recently, the conditions for the separability of the Klein-Gordon equation of the PK metric have also been discussed \cite{1808574}. In the Boyer-Lindquist coordinate system ($t,r,y,\phi$) where $y\equiv\cos\theta$, the contravariant from of the PK metric is given by \cite{Papadopoulos:2018nvd}
\begin{align}
g^{tt}&=\frac{\mathcal{A}_5(r)+\mathcal{B}_5(y)}{\mathcal{A}_1(r)+\mathcal{B}_1(y)}\,,\qquad g^{t\phi}=\frac{\mathcal{A}_4(r)+\mathcal{B}_4(y)}{\mathcal{A}_1(r)+\mathcal{B}_1(y)}\,,\nonumber\\
g^{\phi\phi}&=\frac{\mathcal{A}_3(r)+\mathcal{B}_3(y)}{\mathcal{A}_1(r)+\mathcal{B}_1(y)}\,,\qquad g^{yy}=\frac{\mathcal{B}_2(y)}{\mathcal{A}_1(r)+\mathcal{B}_1(y)}\,,\nonumber\\
g^{rr}&=\frac{\mathcal{A}_2(r)}{\mathcal{A}_1(r)+\mathcal{B}_1(y)}\,,\label{PKmetric}
\end{align}
where $\mathcal{A}_i(r)$ and $\mathcal{B}_i(y)$ are arbitrary functions of $r$ and $y$, respectively. It has been shown that the PK metric \eqref{PKmetric} preserves the separability of the geodesic equations. Actually, several parametrized Kerr-like metrics in the literature, whose geodesic equations are required to be separable, are subclasses of the PK metric. We refer the readers to \cite{Chen:2019jbs} for detailed discussions.

We construct the Kerr-like metric by firstly assuming the radial metric functions to be Kerrian
\begin{align}
\mathcal{A}_1=r^2\,,\quad \mathcal{A}_2=\Delta\,,\quad \mathcal{A}_3=-\frac{a^2}{\Delta}\,,\nonumber\\
\mathcal{A}_4=-\frac{aX}{\Delta}\,,\qquad\mathcal{A}_5=-\frac{X^2}{\Delta}\,,\label{Kerrianmetric}
\end{align} 
where $\Delta\equiv r^2-2Mr+a^2$ and $X\equiv r^2+a^2$. The polar metric functions $\mathcal{B}_i(y)$, on the other hand, are assumed to be \cite{Chen:2020aix}:
\begin{align}
\mathcal{B}_1=a^2y^2+\tilde\epsilon_1(y)\,,\quad \mathcal{B}_2=1-y^2+\tilde\epsilon_2(y)\,,\quad\mathcal{B}_3=\frac{1}{1-y^2}+\tilde\epsilon_3(y)\,,\nonumber\\
\mathcal{B}_4=a+\tilde\epsilon_4(y)\,,\qquad \mathcal{B}_5=a^2(1-y^2)+\tilde\epsilon_5(y)\,,
\end{align}
where $\tilde\epsilon_i(y)$ are arbitrary functions of $y$. These functions $\tilde\epsilon_i(y)$ quantify the deviations from the Kerr spacetime in terms of the polar angle $\theta$. Note that on the above expressions, the parameters $M$ and $a$ stand for the mass and the spin of the black hole, respectively.

In Ref.~\cite{Chen:2020aix}, we have already shown that only the deviation functions $\tilde\epsilon_1$ and $\tilde\epsilon_5$ are allowed to be non-zero in order to satisfy the asymptotic flatness condition. More explicitly, in the asymptotic region, the above metric can be approximated as
\begin{align}
g_{tt}&=-1+\frac{2M}{r}+\mathcal{O}\left(r^{-2}\right)\,,\qquad g_{rr}=1+\frac{2M}{r}+\mathcal{O}\left(r^{-2}\right)\,,\nonumber\\
g_{yy}&=r^2\left[\frac{1}{1-y^2+\tilde\epsilon_2}+\mathcal{O}\left(r^{-2}\right)\right]\,,\qquad g_{\phi\phi}=r^2\left[\frac{1-y^2}{1+\tilde\epsilon_3-y^2\tilde\epsilon_3}+\mathcal{O}\left(r^{-2}\right)\right]\,,\nonumber\\
g_{t\phi}&=\frac{\tilde\epsilon_4}{\frac{1}{1-y^2}+\tilde\epsilon_3}-\frac{2M\left(1-y^2\right)\left(a+\tilde\epsilon_4\right)}{r\left[1+\left(1-y^2\right)\tilde\epsilon_3\right]+\mathcal{O}\left(r^{-2}\right)}\,.
\end{align}
Apparently, one can redefine the coordinate $y$ such that $\tilde\epsilon_2=0$. In addition, we need $\tilde\epsilon_3=\tilde\epsilon_4=0$, otherwise the spacetime is not asymptotically flat. Furthermore, it can be shown that the coefficient of the $1/r^2$ term in the expansion of $g_{tt}$ is proportional to $\tilde\epsilon_1+\tilde\epsilon_5$. More precisely, we have
\begin{equation}
g_{tt}=-1+\frac{2M}{r}-\frac{\tilde\epsilon_1+\tilde\epsilon_5}{r^2}+\mathcal{O}\left(r^{-3}\right)\,.
\end{equation}
Given that the post-Newtonian parameters in the Kerr-like metric is given by
\begin{equation}
\beta=\gamma+\frac{\tilde\epsilon_1+\tilde\epsilon_5}{2M^2}\,,\qquad \gamma=1\,,
\end{equation}
and they have been strongly constrained by the Solar System tests \cite{Williams:2004qba}:
\begin{equation}
|\beta-1|<2.3\times 10^{-4}\,,\qquad |\gamma-1|<2.3\times 10^{-5}\,,
\end{equation}
we will assume that $\tilde\epsilon_1=-\tilde\epsilon_5=\tilde\epsilon$ for the rest of the paper such that the Kerr-like metric satisfies the Solar System tests \cite{Williams:2004qba}. Note that the remaining function $\tilde\epsilon$ cannot be constrained by the Solar System tests. Finally, we will assume that the deviation function $\tilde\epsilon$ is linear in $y$, which turns out to be the simplest choice to break the $\mathbb{Z}_2$ symmetry of the spacetime. In fact, by transforming the metric from the Boyer-Lindquist coordinate system to a so-called ACMC-N
(asymptotically Cartesian and mass centered to order N) coordinates \cite{Thorne:1980ru}, we have shown that the Kerr-like metric considered here has the same gravitational multipoles as the Kerr one, assuming that the Kerr-like metric satisfies the Einstein equations sourced by some artificial matter distributions. This renders our model quite different from that in \cite{Fransen:2022jtw} in which the reflection asymmetry is controlled by non-zero odd
parity multipole moments.

At this point, we would like to emphasize that the deviation function $\tilde\epsilon$ is dimensional. In fact, it has dimension of $[L^2]$ and there is to some extent arbitrariness of choosing the scaling coefficient. For example, in Ref.~\cite{Chen:2020aix}, we simply used the mass squared of the black hole as the coefficient, namely, $\tilde\epsilon(y)=\epsilon M^2y$. However, a direct consequence of this choice is that such a violation of $\mathbb{Z}_2$ symmetry would appear not only in rotating black holes, but also in non-rotating black holes. This may not be a natural choice since it is well-known that in the framework of effective theories, the parity violation and the $\mathbb{Z}_2$ asymmetry usually appear accompanied by some couplings with the dual of the Riemann tensor. This means that physically, they would appear only when the black hole starts spinning. Therefore, in this paper we will take this into account and consider a different choice of the coefficients.

Before closing this section, let us summarize how the deviation functions and the associated parameters are introduced in this model. First of all, we consider the PK metric \eqref{PKmetric}, which is a general stationary and axisymmetric metric allowing for completely separable geodesic equations. The metrics contain five arbitrary functions $\mathcal{A}_i(r)$ of $r$ and five arbitrary functions $\mathcal{B}_i(y)$ of $y$. Since we are focusing on the deviations that induce $\mathbb{Z}_2$ asymmetry in the spacetime, we fix the radial metric functions to be Kerrian (see Eqs.~\eqref{Kerrianmetric}). Then, the polar metric functions $\mathcal{B}_i(y)$ are assumed to be in the Kerrian form \textit{added by some deviation functions} $\tilde\epsilon_i(y)$. These deviation functions should be functions of the polar angle and could induce the $\mathbb{Z}_2$ asymmetry of the spacetime. We also require the spacetime to be asymptotically flat and satisfy the Solar System tests. As a consequence, only one deviation function $\tilde\epsilon(y)$ remains. Later, we will assume that the deviation function $\tilde\epsilon(y)$ is linear in $y$, rendering the multipoles of the spacetime identical as those of Kerr one. As we have mentioned, the deviation function is dimensional. Therefore, we will assign the scaling coefficient and define the associated deviation parameters in a way that the asymmetry is induced by the spin of the black hole, i.e., it appears only when the black hole is spinning.


\section{Null geodesic equations and shadows}\label{secIII}

In this section, we will derive the null geodesic equations of the Kerr-like spacetime constructed in the previous section and investigate the shadow critical curve of such black holes. As we have shown, the $\mathbb{Z}_2$ symmetry of the Kerr-like spacetime is generically broken because of the presence of the deviation function $\tilde\epsilon_1(y)=-\tilde\epsilon_5(y)=\tilde\epsilon(y)$.{\footnote{In Ref.~\cite{Chen:2020aix}, we treated $\tilde\epsilon_1$ and $\tilde\epsilon_5$ as two independent functions and found that the shadow critical curve is completely blind to $\tilde\epsilon_1$.}} We will investigate how the shadow critical curve differs from that of the Kerr black hole in the presence of the deviation function, and how we can place constraints on the deviation function using Sgr A* images. 

\subsection{Null geodesic equations}

The separability of the geodesic equations will be exhibited by using the Hamilton-Jacobi approach. First, we consider the following Lagrangian:
\begin{equation}
\mathcal{L}=\frac{1}{2}g_{\mu\nu}\dot{x}^\mu\dot{x}^\nu\,,
\end{equation}
where the dot denotes the derivative with respect to an affine parameter $\tau$. Since the metric functions are independent of $t$ and $\phi$, there are two constants of motion, namely, the conserved energy $E\equiv-\partial\mathcal{L}/\partial\dot{t}$ and the conserved azimuthal angular momentum $L_z\equiv\partial\mathcal{L}/\partial\dot{\phi}$. One gets the following two geodesic equations
\begin{align}
\Delta\left(r^2+a^2y^2+\tilde\epsilon\right)\dot{t}&=E\left[X^2-\Delta a^2\left(1-y^2\right)+\Delta\tilde\epsilon\right]-2Mra L_z\,,\label{t1}\\
\Delta\left(r^2+a^2y^2+\tilde\epsilon\right)\dot{\phi}&=2Mra E+\left(\frac{r^2-2Mr+a^2y^2}{1-y^2}\right)L_z\,.\label{psi1}
\end{align}

Furthermore, the geodesic equations for $r$ and $y$ are separable and this can be shown by considering the Hamilton-Jacobi equation
\begin{equation}
\frac{\partial\mathcal{S}}{\partial\tau}+\mathcal{H}=0\,,\label{HJeq1}
\end{equation}
where $\mathcal{S}$ is the Jacobi action and $\mathcal{H}$ is the Hamiltonian. The Hamiltonian corresponding to the geodesic Lagrangian can be written as
\begin{equation}
\mathcal{H}=\frac{1}{2}p_\mu p^\mu\,,
\end{equation}
where $p_\mu$ is the conjugate momentum, with which the Hamilton-Jacobi equation \eqref{HJeq1} can be written as 
\begin{equation}
\frac{\partial\mathcal{S}}{\partial\tau}=-\frac{1}{2}g^{\mu\nu}\frac{\partial\mathcal{S}}{\partial x^\mu}\frac{\partial\mathcal{S}}{\partial x^\nu}\,.\label{HJeq2}
\end{equation}
For a separable Hamilton-Jacobi equation, we can consider the following Jacobi action for null geodesics:
\begin{equation}
\mathcal{S}=-Et+L_z\phi+\mathcal{S}_r(r)+\mathcal{S}_y(y)\,.\label{ansztzS}
\end{equation}
Inserting the ansatz \eqref{ansztzS} intro the Hamilton-Jacobi equation \eqref{HJeq2}, one gets
\begin{align}
-\frac{X^2}{\Delta}E^2&+\frac{2aX}{\Delta}EL_z-\frac{a^2}{\Delta}L_z^2+\Delta\left(\frac{d\mathcal{S}_r}{dr}\right)^2\nonumber\\
&=\left[-a^2\left(1-y^2\right)+\tilde\epsilon\right]E^2+2aEL_z-\frac{L_z^2}{1-y^2}-\left(1-y^2\right)\left(\frac{d\mathcal{S}_y}{dy}\right)\,.\label{sep}
\end{align}
The left-hand side of the above equation \eqref{sep} depends only on $r$, while the right-hand side depends only on $y$. One can then introduce a decoupling constant $K$ such that the equation can be separated into
\begin{align}
\left(r^2+a^2y^2+\tilde\epsilon\right)\dot{r}&=\pm \sqrt{E^2R(r)}\,,\label{geo1}\\
\left(r^2+a^2y^2+\tilde\epsilon\right)\dot{\theta}&=\pm \sqrt{E^2\Theta(\theta)}\,,\label{geo2}
\end{align}
where
\begin{align}
R(r)&\equiv\left(r^2+a^2-a\xi\right)^2-\Delta\left[\eta+\left(\xi-a\right)^2\right]\,,\label{Rr}\\
\Theta(\theta)&\equiv\eta+\cos^2\theta\left(a^2-\xi^2\csc^2\theta\right)+\tilde\epsilon\,.
\end{align}
Note that we have defined $\xi\equiv L_z/E$ and $\eta\equiv K/E^2$ in the above equations. Expressed with these parameters, Eqs.~\eqref{t1} and \eqref{psi1} can be rewritten as
\begin{align}
\Delta\left(r^2+a^2y^2+\tilde\epsilon\right)\dot{t}/E&=\left[X^2-\Delta a^2\left(1-y^2\right)+\Delta\tilde\epsilon\right]-2Mra\xi\,,\label{t2}\\
\Delta\left(r^2+a^2y^2+\tilde\epsilon\right)\dot{\phi}/E&=2Mra+\left(\frac{r^2-2Mr+a^2y^2}{1-y^2}\right)\xi\,.\label{psi2}
\end{align}

\subsection{Shadow critical curve}

Here we will derive the equations that describe the shadow critical curve. Essentially, the critical curve is the impact parameter of the photon region around the black hole \cite{Bardeen1973}. The photon region is composed of several spherical photon orbits, each with its own radius $r_p$. The photon moving on a spherical photon orbit with radius $r_p$ has its own constants of motion $\xi$ and $\eta$, which can be determined by $r_p$. The spherical orbits are characterized by the following conditions
\begin{equation}
R(r_p)=\frac{dR}{dr}\Bigg|_{r_p}=0\qquad \frac{d^2R}{dr^2}\Bigg|_{r_p}\ge0\,,
\end{equation}
which, after using Eq.~\eqref{Rr}, give (see Ref.~\cite{Bardeen1973} for detailed discussions)
\begin{align}
\xi(r_p)&=\frac{1}{a}\left(r^2+a^2-\frac{4\Delta r}{\Delta'}\right)\Bigg|_{r=r_p}\,,\label{xirp}\\
\eta(r_p)&=\frac{16r^2\Delta}{\Delta'^2}\Bigg|_{r=r_p}-\left[\xi(r_p)-a\right]^2\,,\label{etarp}
\end{align}
where the prime denotes the derivative with respect to $r$. Therefore, it can be seen that for a photon moving on a spherical orbit, its associated constants of motion $\xi$ and $\eta$ are determined directly by the orbital radius $r_p$ according to Eqs.~\eqref{xirp} and \eqref{etarp}. It should be also noted that $\Theta$ can not be negative for a viable photon orbit. The zeros of $\Theta$ determines the extreme latitude $\theta_m$ on a given spherical orbit. If the deviation function $\tilde\epsilon$ is an even function of $y$, the spherical motion of photons is $\mathbb{Z}_2$ symmetric. However, if $\tilde\epsilon$ is not an even function of $y$, the $\mathbb{Z}_2$ symmetry could be broken and the extreme latitudes on the north and the south hemispheres could be different. For a complete discussion of the spherical photon orbits around a Kerr black hole, we would like to refer the readers to the paper \cite{Teo:2003} and the review \cite{Perlick:2010zh}. Very recently, the spherical orbits of massive particles moving around a Kerr black hole have also been discussed in Ref.~\cite{Teo:2020sey}.

\begin{figure}
\centering
\graphicspath{{fig/}}
\includegraphics[scale=0.4]{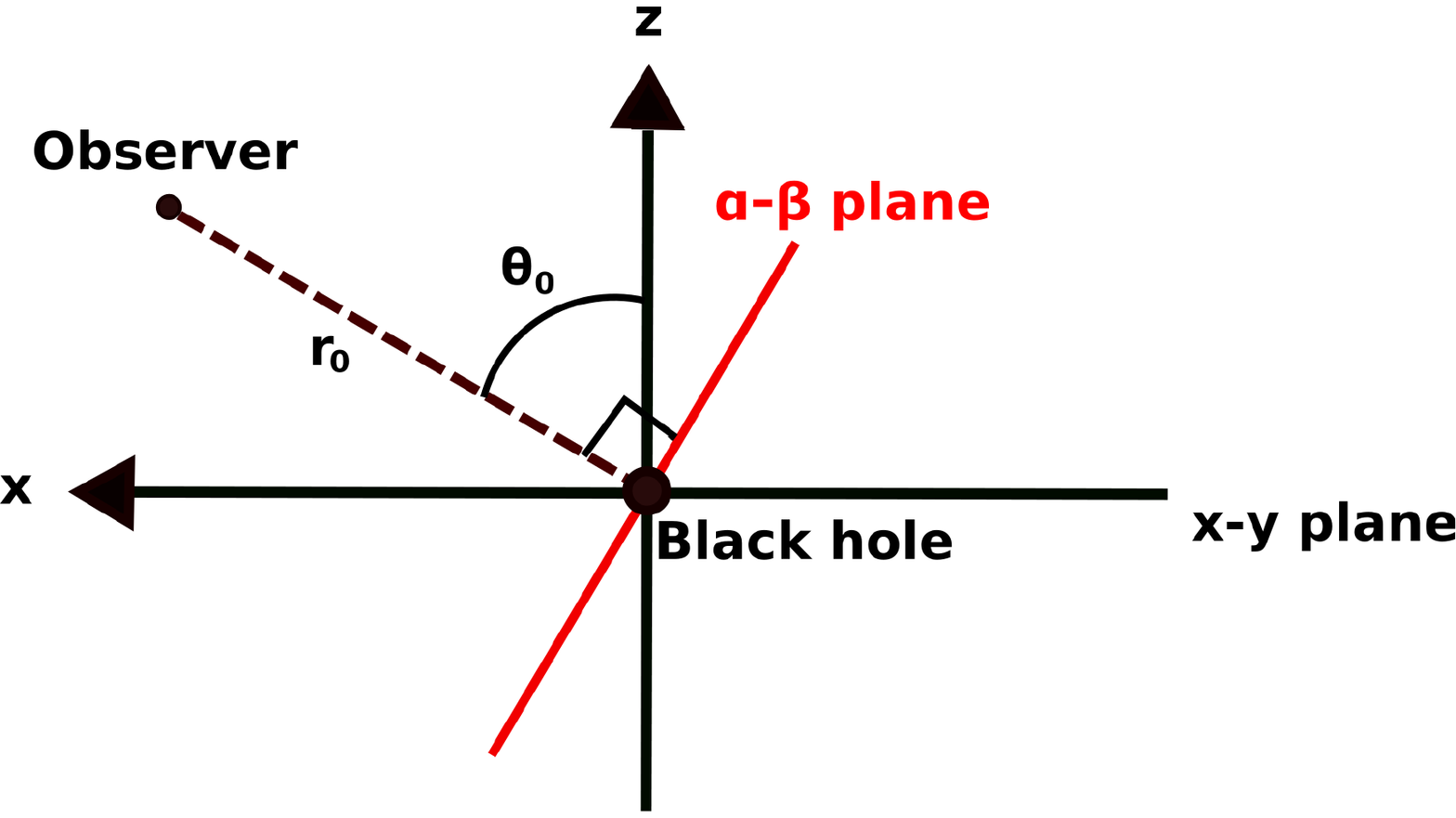}
\caption{The schematic plot of the celestial coordinate system. The black hole is located at the origin and the observer is assumed to be on the $x$-$z$ plane. The $z$-axis corresponds to the axis of symmetry of the black hole. The position of the observer can also be expressed with spherical coordinates ($r_0,\theta_0,\phi_0$) in which $\phi_0$ is set to zero. The black hole shadow is projected on the $\alpha$-$\beta$ plane.} 
\label{figccele}
\end{figure}

The visualization of the apparent shape of a shadow usually relies on the celestial coordinates $(\alpha,\beta)$. We illustrate the definition of the celestial coordinate system in Figure~\ref{figccele}. The coordinate $\alpha$ is the apparent perpendicular distance between the edge of the shadow and the axis of symmetry ($z$-axis). The coordinate $\beta$, on the other hand, is defined by the apparent perpendicular distance between the edge of the shadow and the $y$-axis. Since the spacetime under consideration is asymptotically flat, the celestial coordinates can be expressed as \cite{Vazquez:2003zm}
\begin{equation}
\alpha=\lim_{r_0\rightarrow\infty}\left(-r_0^2\sin\theta_0\frac{d\phi}{dr}\Bigg|_{r_0,\theta_0}\right)\,,\qquad \beta=\lim_{r_0\rightarrow\infty}\left(r_0^2\frac{d\theta}{dr}\Bigg|_{r_0,\theta_0}\right)\,,\label{alphabetadef}
\end{equation}
where $r_0$ is the distance between the observer and the black hole, and $\theta_0$ defines the inclination angle between the rotation axis of the black hole and the line of sight of the observer. Using the geodesic equations \eqref{geo1}, \eqref{geo2}, \eqref{t2}, and \eqref{psi2}, one can derive the celestial coordinates of the shadow critical curve as follows:
\begin{equation}
\alpha=-\frac{\xi}{\sin\theta_0}\,,\qquad \beta=\pm\sqrt{\eta+a^2\cos^2\theta_0-\xi^2\cot^2\theta_0+\tilde\epsilon(y_0)}\,,\label{alphabeta}
\end{equation}
where $y_0\equiv\cos\theta_0$. Essentially, the critical curve of a black hole shadow can be obtained by firstly treating $r_p$ as a running variable to parametrize $\xi(r_p)$ and $\eta(r_p)$, then inserting $\xi(r_p)$ and $\eta(r_p)$ into Eq.~\eqref{alphabeta}. The points on the contour with different values of $\alpha$ are contributed by the photons moving on different spherical photon orbits. Unlike the case of non-rotating black holes which only have a single spherical photon orbit (the photon sphere), a rotating black hole is surrounded by multiple spherical photon orbits and this collection of orbits is called photon region. As we have mentioned, the shadow critical curve of a rotating black hole is contributed by the photons moving on the spherical orbits in the photon region. 

It should be emphasized at this point that although the spacetime is not $\mathbb{Z}_2$ symmetric, the critical curve is symmetric with respect to the $\alpha$-axis, irrespective of the inclination angle $\theta_0$. This important result has been pointed out in Refs.~\cite{Cunha:2018uzc,Grenzebach:2014fha,Lima:2021las} and can be seen directly from the second equation of \eqref{alphabeta}. As has been discussed in Ref.~\cite{Cunha:2018uzc}, the reason for the preservation of this symmetry of the critical curve strongly depends on the separability of the geodesic equations. More explicitly, the separability of the Hamilton-Jacobi equation already implies that in the $\theta$ sector of the geodesic equations, the 4-momentum $p_\theta$ always appears in the form of $p_\theta^2$. According to Eq.~\eqref{alphabetadef}, the coordinate $\beta$ is proportional to $p_\theta$, therefore, the critical curve is symmetric with respect to the $\alpha$-axis. In Ref.~\cite{Chen:2020aix}, we have demonstrated that the light rays contributing to two points slightly away from the shadow critical curve with celestial coordinates ($\alpha_p,\pm\beta_p$) are corresponding to the same spherical photon orbit, although the detailed motions of the photons on that orbit are $\mathbb{Z}_2$ asymmetric (see Figure 5 in Ref.~\cite{Chen:2020aix}).


\section{Results and constraints from Sgr A* images}\label{sec.IV}
In this section, we will present how the shadow critical curves, particularly, the size of the contours, are affected by the deviation function $\tilde\epsilon(y)$. As we have mentioned, even though we have assumed $\tilde\epsilon(y)$ to be linear in $y$, the deviation function $\tilde\epsilon(y)$ is not dimensionless and its explicit dependence on the mass and the spin of the black hole is not unique. For example, in Ref.~\cite{Chen:2020aix} we assumed $\tilde\epsilon(y)=\epsilon M^2y$. However, this choice may not be natural from the physical point of view because the $\mathbb{Z}_2$ symmetry is broken as well for non-rotating black holes in this case. If we require that the violation of $\mathbb{Z}_2$ symmetry is induced in the presence of spin \cite{Cardoso:2018ptl,Cano:2019ore}, it is natural to consider the following choice of the deviation function: $\tilde\epsilon(y)=\epsilon aMy$. Note that this deviation function has been chosen in \cite{Chen:2021ryb} to investigate the timelike circular orbits and accretion disk properties of the Kerr-like metric.

\begin{figure}
\centering
\graphicspath{{fig/}}
\includegraphics[scale=0.5]{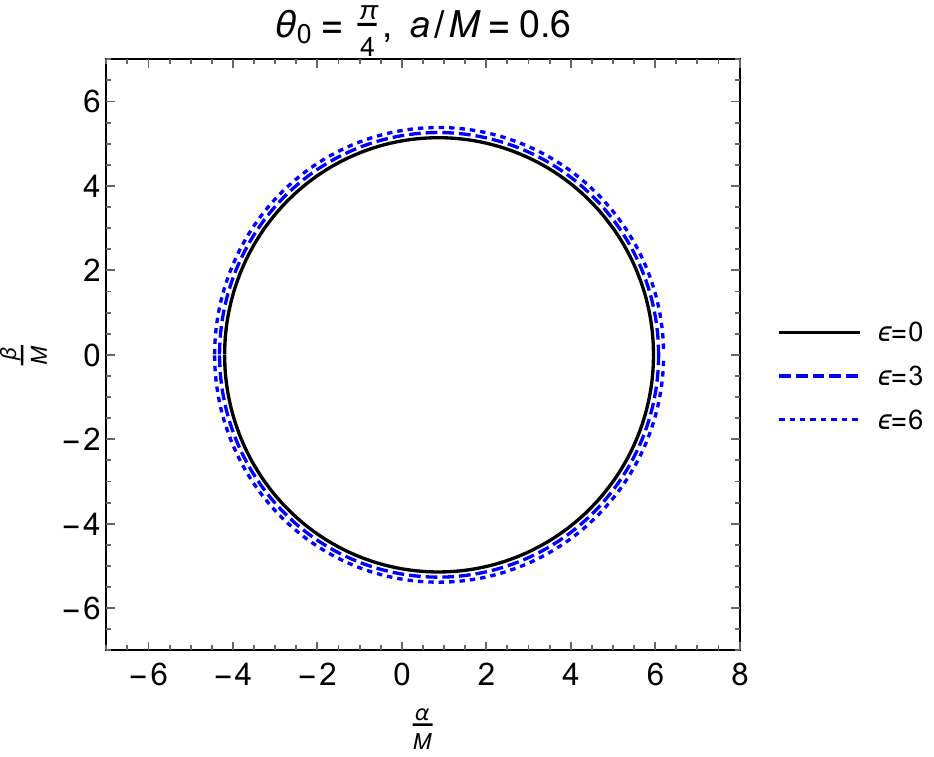}
\includegraphics[scale=0.5]{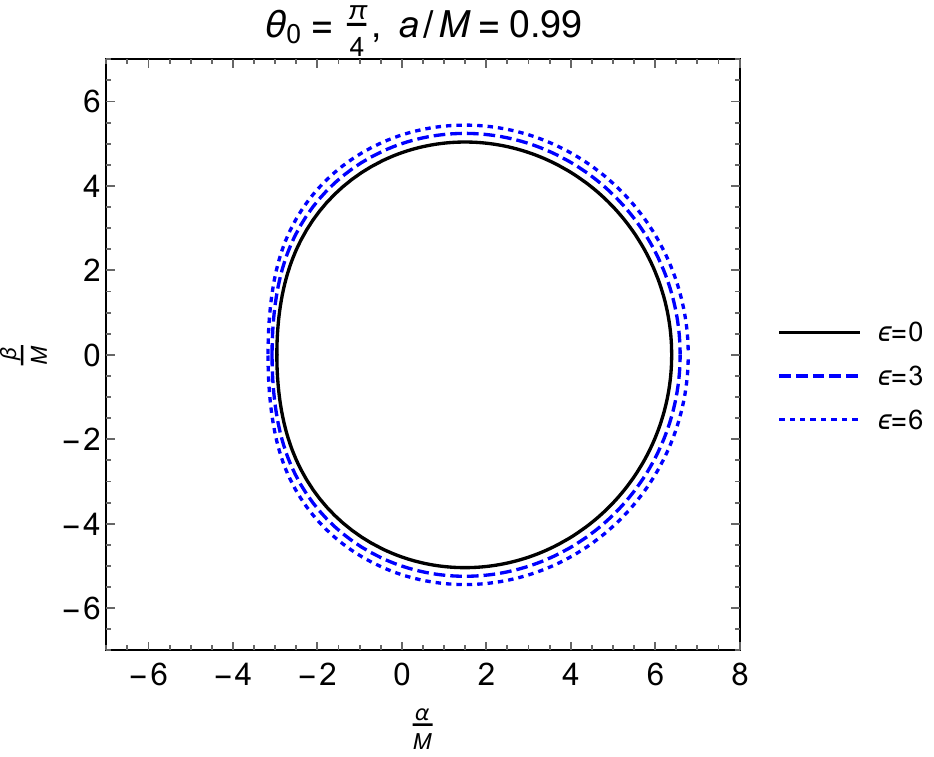}
\caption{The critical curves cast by the Kerr-like black hole with $\tilde\epsilon(y)=\epsilon aMy$ with different values of the dimensionless parameter $\epsilon$. The left (right) panel shows the results with $a/M=0.6$ ($a/M=0.99$). The inclination angle is assumed to be $\pi/4$. The black curves represent the critical curves cast by Kerr black holes.} 
\label{figshadowcon1}
\end{figure}

In Figure~\ref{figshadowcon1}, we fix the inclination angle ($\theta_0=\pi/4$) and show the critical curves with $a/M=0.6$ (left) and $a/M=0.99$ (right) with respect to different values of $\epsilon$. When $\theta_0=\pi/4$, one can see that increasing $\epsilon$ would increase the apparent size of the contour. We also find that for edge-on observers ($\theta_0=\pi/2$), the critical curve appears to be indistinguishable from its Kerr counterpart since the deviation function vanishes. On the other hand, for face-on observers ($\theta_0=0$ or $\pi$), the critical curve becomes a perfect circle, as expected, whose radius is altered by the deviation parameter $\epsilon$.

Having discussed how the deviation parameter $\epsilon$ could affect the size of the critical curve, we will place constraints on the parameter space using the Sgr A* images released by the Event Horizon Telescope (EHT) Collaboration. In fact, constraining non-Kerr parameters is already possible using the ring size of M87* and has been widely considered in the literature \cite{Psaltis:2020lvx,Brahma:2020eos,EventHorizonTelescope:2021dqv,Meng:2022kjs}. For the case of M87*, the image size is consistent with GR's predictions for critical curves within $17\%$ \cite{Akiyama:2019eap}. As for the Sgr A* images, the extensive studies of stellar orbits around it allows accurate measurements of the mass and distance of the black hole \cite{Do:2019txf,GRAVITY:2021xju}. In terms of the fractional diameter deviation $\delta$, which is defined by the relative difference between the analytic
shadow diameter and the Schwarzschild diameter:
\begin{equation}
\delta=\frac{\bar{d}}{6\sqrt{3}}-1\,,
\end{equation}
where $\bar{d}$ is the median diameter of the critical curve, we have the constraints \cite{EventHorizonTelescope:2022xqj}
\begin{equation}
\delta=-0.04^{+0.09}_{-0.10}\quad\textrm{and}\quad\delta=-0.08^{+0.09}_{-0.09}\,,\label{boundsdelta}
\end{equation}
which are based on the
Keck and VLTI priors on the mass-to-distance ratio, respectively.

\begin{figure}
\centering
\graphicspath{{fig/}}
\includegraphics[scale=0.5]{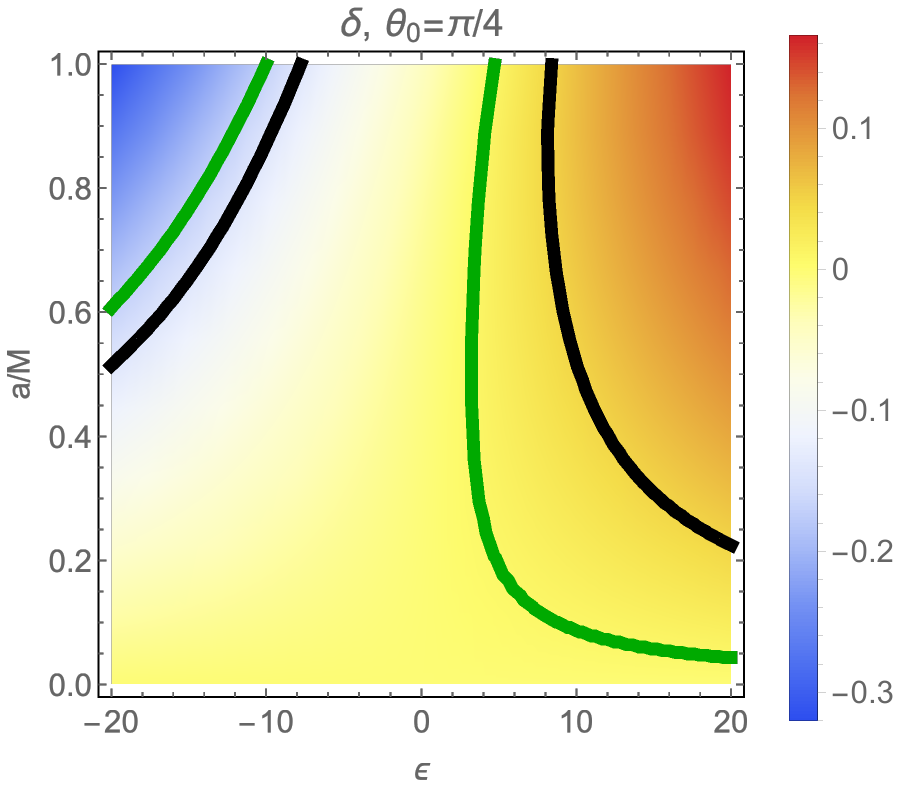}
\includegraphics[scale=0.5]{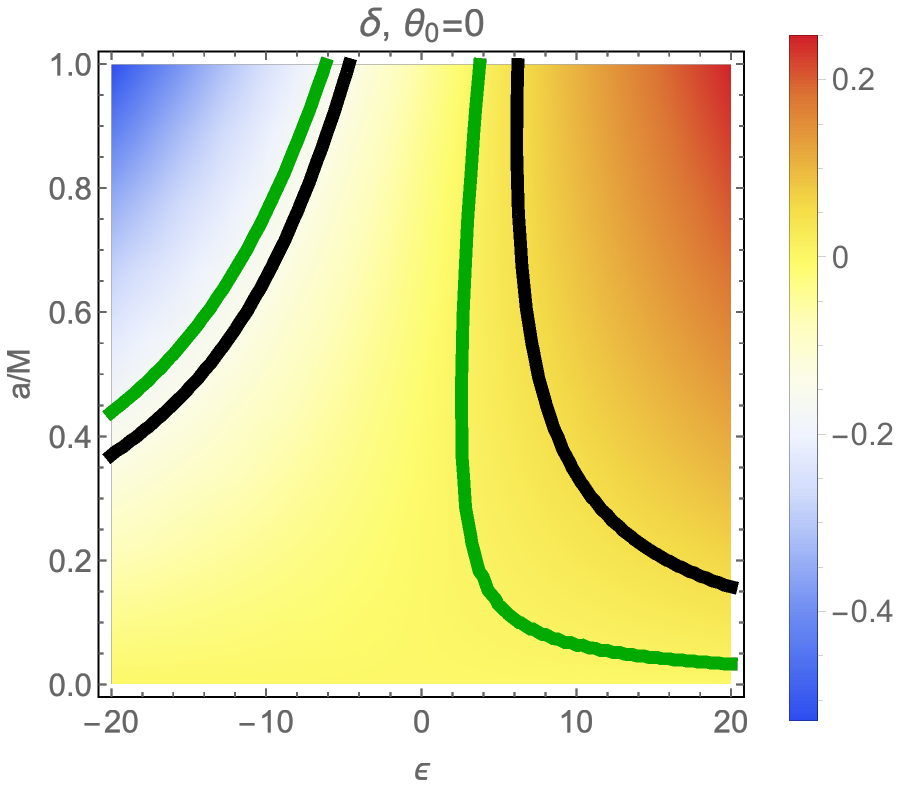}
\caption{These density profiles show how the fractional diameter deviation $\delta$ predicted by the Kerr-like metric is changed with respect to $\epsilon$ and the spin $a/M$. The inclination angle is chosen to be $\theta_0=\pi/4$ (left) and $\theta_0=0$ (right), respectively. The black curves and green curves indicate the bounds that are calculated using the Keck and VLTI priors on the mass-to-distance ratio, i.e., Eq.~\eqref{boundsdelta}. The allowed parameter space is lying between a pair of each colored curves.} 
\label{fig.thetaconstraint}
\end{figure}

We then calculate the fractional diameter deviation $\delta$ for the Kerr-like metric in the parameter space $\{a,\epsilon\}$. The results are shown in Figure~\ref{fig.thetaconstraint}. In the left panel, we choose the inclination angle $\theta_0=\pi/4$, while on the right panel, we choose $\theta_0=0$. Note that the current images of EHT can rule out the scenario where the black hole is viewed at high inclination ($\theta_0>50^{\circ}$) \cite{EventHorizonTelescope:2022xnr}. A non-spinning black hole scenario has also been excluded \cite{EventHorizonTelescope:2022xnr}.  In Figure~\ref{fig.thetaconstraint}, the black and green curves indicate the bounds that are calculated using the Keck and VLTI priors on the mass-to-distance ratio, i.e., Eq.~\eqref{boundsdelta}. 

One can see that the parameter space can be constrained, in particular, when the black hole is rapidly spinning. If the black hole is slowly spinning, constraints on $\epsilon$ are very weak. In addition, a tighter constraint on the parameter space can be placed if the inclination angle is smaller, with the tightest constraints being placed in the face-on scenario.

\section{Conclusions}\label{conclu}
In this paper, we adopt a theory-agnostic approach and consider a class of Kerr-like spacetimes whose $\mathbb{Z}_2$ symmetry is generically broken. The spacetime is asymptotically flat and the geodesic equations are assumed to be completely separable. Based on these assumptions, the $\mathbb{Z}_2$ asymmetry of the spacetime can be described by one arbitrary function of polar angle $\theta$. Rotating black holes without $\mathbb{Z}_2$ symmetry usually appear in gravitational theories which contain parity-violating terms in the equations of motion, or some effective low-energy theories of a fundamental quantum theory of gravity. The Kerr-like spacetime that we construct in this paper can serve as an approximation to the rotating black holes of such complicated gravitational theories.

As opposed to the original work \cite{Chen:2020aix} in which the deviation function does not depend on spin, in this paper we consider the deviation function in a way that the $\mathbb{Z}_2$ asymmetry is induced by the spin of the black hole. This assumption is natural because in most effective theories of quantum gravity, the parity-violating terms are accompanied by the dual of Riemann tensor, and these terms are usually zero in a spherically symmetric spacetime.

After constructing the Kerr-like metric, we investigate the shadow critical curve cast by the Kerr-like black holes. We confirm once again that although the spacetime is not $\mathbb{Z}_2$ symmetric, the shadow contour is always symmetric with respect to the horizontal axis. As we have mentioned, this symmetry is related to the separability of the geodesic equations. Then, we investigate how the deviation function changes the apparent size of the critical curve. It turns out that the deviation parameter $\epsilon$ can be constrained as long as the mass of the black hole and its distance from the earth can be measured accurately using other independent methods. As we have illustrated in Figure~\ref{fig.thetaconstraint}, the latest bounds on the ring size of Sgr A* images can place tighter constraints on the parameter space than those put from the M87* image, giving rise to an important test on equatorial reflection symmetry of black holes.

Constraining non-GR models using shadow critical curves is certainly a too ideal scenario. A realistic black hole image consists of images from the accretion disk as well. It has been shown in \cite{Chen:2021ryb} that the violation of equatorial reflection symmetry renders the accretion disk a curved surface. A reliable thin-disk model describing a curved disk is needed to incorporate the disk image into the full context of the image. Furthermore, although the critical curve remains vertically symmetric, the $n<\infty$ photon rings \cite{Johnson:2019ljv,Gralla:2019drh}, where $n$ is the winding number of photons rotating around the black hole before being observed, could have vertically asymmetric distribution for an edge-on observer. These photon rings are important observational targets for next generation EHT (ngEHT) collaboration. We will tackle these interesting topics in the future.

\section*{Acknowledgement}

CYC is supported by the Institute of Physics of Academia Sinica.

\end{document}